\documentclass[conference]{IEEEtran}
\IEEEoverridecommandlockouts
\usepackage{cite}
\usepackage{amsmath,amssymb,amsfonts}
\usepackage{graphicx}
\usepackage{textcomp}
\usepackage{xcolor}
\usepackage{url}
\def\BibTeX{{\rm B\kern-.05em{\sc i\kern-.025em b}\kern-.08em
    T\kern-.1667em\lower.7ex\hbox{E}\kern-.125emX}}
\begin{document}

\title{Open-Source LLM-Driven Formal Verification:\\A Multi-Agent Pipeline for RTL Repair}

\author{\IEEEauthorblockN{Ha Trung Tran}
\IEEEauthorblockA{\textit{Independent Researcher} \\
Seattle, WA, USA \\
trantrungha072020@gmail.com}
}

\maketitle

\begin{abstract}
Verification consumes the majority of modern chip design effort, yet the formal verification tools that provide mathematical guarantees of correctness remain expensive and restrictively licensed. While large language models (LLMs) have shown promise for hardware design, existing approaches to RTL repair validate their results through simulation---which exercises only a subset of inputs---or rely on commercial tools, and few combine formal proof with an entirely open-source toolchain. In this paper, we present a multi-agent pipeline that couples an LLM with an open-source formal backend (Yosys, SymbiYosys, and Z3) to repair RTL through counterexample-guided iteration: the framework generates formal properties, verifies the design, and feeds counterexamples back to the LLM until the design is proved correct by k-induction or an iteration budget is exhausted. Through an ALU case study, we show that the pipeline can detect and repair a real functional bug with a formal proof of correctness. Across a six-benchmark suite, one design is repaired reliably, and we characterize four distinct failure modes---bounded-cover vacuity, specification ambiguity, temporal-logic bugs, and multi-property pressure. We frame this work as a feasibility study with a detailed failure analysis, and additionally report a practical limitation of the Yosys bind directive relevant to the open-source formal verification community.
\end{abstract}


\section{Introduction}
Verification consumes the majority of modern chip design effort, often accounting for 60--70\% of the total design cycle. Formal verification, which provides mathematical guarantees of correctness, is a powerful tool for this task, yet the industrial-grade tools that perform it remain expensive and restrictively licensed. In parallel, large language models (LLMs) have emerged as a promising means of automating parts of the hardware design flow, from generating RTL to assisting with debugging.

Existing work at this intersection, however, leaves a gap. LLM-based approaches to RTL either generate designs and validate them through simulation---which exercises only a subset of the input space---or rely on commercial verification tools. Comparatively few combine formal proof, which reasons over all reachable states, with an entirely open-source toolchain that is reproducible and freely accessible.

In this paper, we present a multi-agent pipeline that couples an LLM with an open-source formal backend (Yosys, SymbiYosys, and Z3) to repair RTL through counterexample-guided iteration. The pipeline generates formal properties from a specification, verifies the design, and---when verification fails---feeds the resulting counterexample back to the LLM to guide a repair, repeating until the design is either proved correct or the iteration budget is exhausted. Because correctness is established by formal proof rather than simulation, a successful repair is guaranteed to hold for all reachable inputs.

We evaluate the framework on a suite of six benchmarks. Through an ALU case study, we show that the pipeline can detect and repair a real functional bug and prove the result correct by k-induction, demonstrating the feasibility of the approach. Across the full suite, however, one of six benchmarks is repaired reliably, and we characterize four distinct failure modes on the remainder: bounded-cover vacuity, specification ambiguity, temporal-logic bugs, and multi-property pressure. We therefore frame this work as a feasibility study with a detailed failure analysis; the clear boundary it draws between tractable and intractable cases is itself an informative result.

This paper makes the following contributions:
\begin{itemize}
\item A multi-agent framework for LLM-driven RTL repair that relies exclusively on open-source formal verification tools.
\item A typed Property IR that separates property reasoning by the LLM from the generation of syntactically correct SystemVerilog assertions.
\item An end-to-end ALU case study together with a characterization of four distinct failure modes, obtained through a multi-run evaluation protocol.
\item A practical engineering note documenting a limitation of the Yosys \texttt{bind} directive relevant to the open-source formal verification community.
\end{itemize}

\section{Background}

\subsection{Formal Verification and Model Checking}
Formal verification uses mathematical reasoning to prove that a design satisfies a given property across all reachable states and inputs, in contrast to simulation, which exercises only specific inputs \cite{biere1999}. Bounded Model Checking (BMC) verifies a property within a bound of $k$ cycles: if a counterexample is found, an error is reported; if none is found within $k$ steps, no conclusion can be drawn for deeper behavior \cite{biere1999}. This bounded nature is the root cause of the bounded-cover vacuity discussed in Section~\ref{sec:discussion}. To prove a property at all depths rather than only within $k$ steps, we rely on k-induction, which establishes a base case and an inductive step \cite{sheeran2000}. This is what allows a PASS in our framework to be interpreted as proven for all reachable inputs. Properties are expressed as SystemVerilog Assertions (SVA); our framework uses immediate assertions. When a property fails, the solver returns a counterexample (CEX) as a waveform trace (VCD) that demonstrates the violation. This trace is the feedback our pipeline consumes, parsed by the CEX Analyzer node.

\subsection{The Open-Source Formal Toolchain}
Our verification backend is built entirely from open-source tools. Yosys is an open-source RTL synthesis framework that reads and elaborates SystemVerilog \cite{yosys}. SymbiYosys (SBY) is a front-end that coordinates the formal flow over Yosys and supports both prove and cover modes \cite{sby}. Z3 serves as the SMT solver backend for bounded model checking and k-induction \cite{z3}. We deliberately adopt this open-source stack because commercial formal tools, such as VCS and JasperGold, are expensive and restrictively licensed, which limits reproducibility and accessibility---particularly for academic labs and individual developers.

\subsection{LLMs for RTL Generation and Repair}
Large language models can generate Verilog from natural-language specifications \cite{verigen,verilogeval}. However, generated code frequently contains both syntax and logic errors, which motivates iterative repair. Existing approaches, such as RTLFixer, combine simulation or compiler feedback with iterative prompting to correct these errors \cite{rtlfixer}. Most such methods rely on simulation, which checks only a limited set of inputs, or on commercial tools. Comparatively few combine formal proof with an entirely open-source toolchain---the gap this work addresses.

\section{Architecture}

\subsection{Overview}
\begin{figure*}[t]
\centering
\includegraphics[width=\columnwidth]{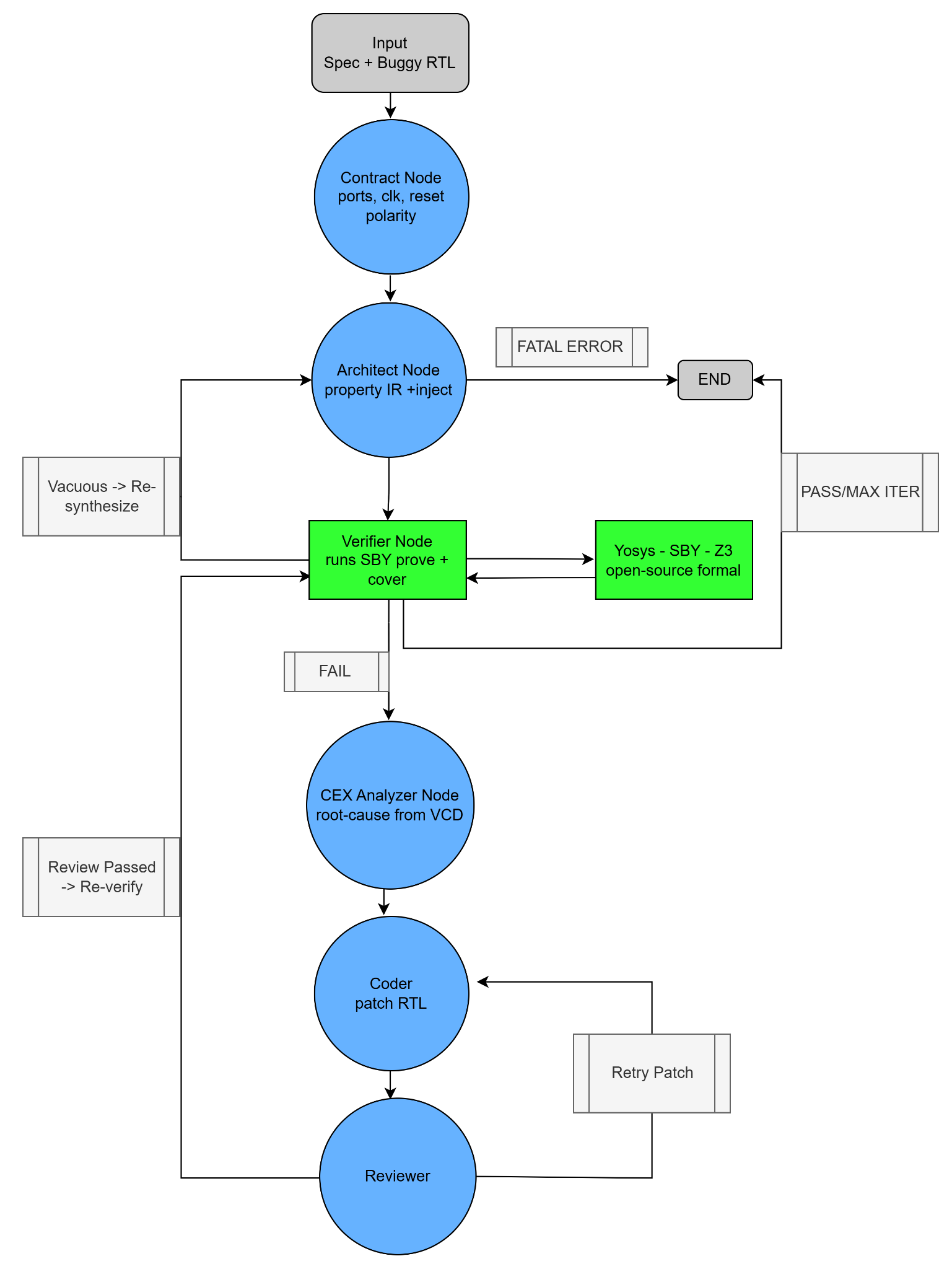}
\caption{System architecture of the multi-agent RTL repair pipeline. Blue nodes are LLM agents; green nodes are formal tools.}
\label{fig:arch}
\end{figure*}

Our framework is a closed-loop pipeline that repairs RTL by combining LLM-based reasoning with formal verification feedback. Rather than tasking a single model with the entire repair problem, we decompose it into a set of specialized agents, each responsible for a narrow, well-defined step: extracting the design interface, synthesizing formal properties, verifying the design, analyzing counterexamples, repairing the RTL, and reviewing the result. Fig.~\ref{fig:arch} shows the overall architecture. The pipeline iterates a verify--repair loop, guided at each step by concrete counterexamples from the formal backend, until the design is proved correct or an iteration budget is exhausted.

\subsection{Design Principles}
The architecture rests on three principles. First, \emph{separation of concerns}: decomposing the pipeline into specialized agents keeps each step controllable and independently debuggable, in contrast to a single monolithic prompt whose failures are difficult to localize. Second, a \emph{formal feedback loop}: unlike approaches that rely on simulation, our pipeline consumes counterexamples---concrete waveform traces that a solver produces to witness a violation---as its repair signal, giving the model precise, grounded evidence of what went wrong. Third, a \emph{separation of reasoning from syntax}: the LLM reasons about \emph{what} properties should hold, expressed through a typed intermediate representation, while a dedicated compiler guarantees that the resulting SystemVerilog assertions are syntactically correct.

\subsection{Pipeline Flow}
The pipeline takes a natural-language specification and a buggy RTL module as input. The Contract agent first extracts the structural interface, which the Architect uses to synthesize formal properties that are compiled and inlined into the design. The Verifier then checks the design and routes the outcome: a PASS terminates the loop; a FAIL forwards the counterexample to the CEX Analyzer, which localizes the fault for the Coder to repair, after which the Reviewer audits the fix before returning control to the Verifier; a vacuous result routes back to the Architect to regenerate properties. This loop repeats until the design is proved correct or the maximum number of iterations is reached.

\section{Implementation}

\subsection{System Overview}
We implement our framework in Python 3.12 as an automated RTL repair system that combines an LLM-driven multi-agent pipeline with open-source formal methods. The pipeline is orchestrated by a state graph (LangGraph 1.0.6) that manages a shared state and performs conditional routing between nodes. The pipeline comprises two kinds of nodes: five nodes are driven by a large language model (GPT-4o) for reasoning, and one node is a tool wrapper that invokes the formal backend. All verification runs on open-source tools (Yosys, SymbiYosys, and Z3), removing any dependency on commercial licenses.

\subsection{Technology Stack}
The system is built on five core components. LangGraph 1.0.6 provides the orchestration layer: it defines each node as part of a graph, routes the shared state according to runtime conditions, and supports the feedback loop between the verifier, coder, and reviewer. GPT-4o, accessed through the OpenAI API (client library 2.15.0), serves as the LLM for the reasoning nodes. Yosys 0.61+39, through its built-in SystemVerilog frontend, reads the RTL together with the inlined assertions and elaborates them into a netlist for the solver. SymbiYosys (SBY) 0.61 orchestrates the formal flow in two modes: prove, which uses k-induction, and cover, which checks reachability. Z3 4.15.5 serves as the SMT backend for bounded model checking and k-induction. This entirely open-source stack ensures that our results are reproducible without any commercial license.

\subsection{Pipeline Nodes}
The pipeline consists of six nodes. The entry node, the Contract node, ingests a natural-language specification together with the RTL module. It extracts the structural interface---ports, clock, reset, and reset polarity---and returns a design contract in JSON form that is shared by all subsequent nodes.

The Architect node receives this design contract and uses the LLM to generate a Property IR, a set of typed objects. An internal Property Compiler then translates the IR into SystemVerilog immediate assertions and injects them inline into the design under test (DUT), returning the merged RTL module.

The Verifier node receives the assertion-inlined RTL. It generates the SBY configuration file and runs Yosys, SBY, and Z3 in both prove and cover modes. Its outcome is one of three states: PASS, FAIL with a VCD counterexample, or VACUOUS.

When the outcome is FAIL with a counterexample, the CEX Analyzer node parses the VCD trace, identifies which assertion is violated, and produces a root-cause analysis together with a repair hint. This yields a textual repair strategy for the Coder node.

The Coder node receives the current RTL module and the repair strategy. The LLM corrects the faulty logic while preserving the inlined assertion block unchanged, and returns the repaired RTL.

The repaired module is then passed to the Reviewer node, which audits the semantic alignment of the fix. If the review passes, control returns to the Verifier node to re-check the updated RTL; otherwise, control returns to the Coder node for a further attempt.

The entire pipeline is orchestrated by the state graph, beginning at the Contract node. The Verifier node is the central branching point, with four possible transitions (to the coder, to the analyzer, back to the architect on vacuity, or to termination on pass or maximum iterations), while the Reviewer node forms a second branching point with two transitions. Fig.~\ref{fig:arch} illustrates the full architecture.

\subsection{Key Engineering Decisions}
\textbf{Typed Property IR.} Rather than having the LLM emit SystemVerilog assertions directly---which is error-prone and difficult to control---we introduce a typed intermediate representation. The LLM generates IR objects drawn from seven types: Equality, Implication, ResetInvariant, MutualExclusion, TransitionProperty, HandshakeProperty, and HoldProperty. The compiler then translates these into valid assertions. This separates property reasoning, performed by the LLM, from syntactically correct generation, guaranteed by the compiler, and allows each IR object to be validated before compilation.

\textbf{Inline Assertion Injection.} Assertions are injected directly into the DUT module, immediately before the final \texttt{endmodule}, rather than through a \texttt{bind} directive. We adopt inline injection because of a specific limitation in the Yosys frontend; the underlying issue that motivates this choice is detailed in Section~\ref{sec:yosysbind}.

\textbf{Robustness Layer.} Because the LLM emits bit-width specifiers inconsistently, we normalize them to a safe form. We further mask arithmetic results to the exact bit-width of the target signal so that assertions match the modular wrap-around behavior of the RTL. Finally, injection is made idempotent by stripping any pre-existing assertion block before re-injecting, which prevents duplicate assertion cells during repeated iterations. Together, these steps ensure that LLM-generated assertions compile reliably through Yosys.

\section{Evaluation}

\subsection{Experimental Setup}
We evaluate whether an LLM, guided by counterexample traces and backed entirely by open-source formal tools, can repair RTL designs without any commercial license. Our setup is defined by five conditions. First, the language model: we use GPT-4o (OpenAI) with the temperature set to zero and a maximum of ten iterations per benchmark. Second, the formal backend: Yosys 0.61+39, SymbiYosys 0.61, and Z3 4.15.5. Third, the benchmark suite: six modules---counter, alu, arbiter, axi\_lite\_slave, uart\_tx, and fifo---ranging from a simple combinational design (alu) to sequential designs with more complex state (fifo, axi\_lite\_slave). Fourth, each benchmark is injected with a single, specific bug. Finally, we run each benchmark five times independently, yielding 30 runs in total.

We use five independent runs per benchmark because the language model is nondeterministic: the same input can produce different outputs across runs. Measuring this variance, rather than reporting a single run, provides the statistical rigor needed to assess the framework's stability.

\subsection{Evaluation Metrics}
We report three metrics. The pass rate is the number of successful runs over the total number of runs for a given benchmark (for example, 5/5 or 0/5). We define a run as PASS only when the repaired RTL satisfies two conditions: every assertion is proved by k-induction, and every cover property is reachable (that is, the result is not vacuous). The iterations to convergence is the number of verifier loops executed before a PASS, or the maximum of ten if the run does not converge. Finally, we record the wall-clock time of each run.

A key point is that our PASS criterion is based on formal proof rather than simulation. Simulation exercises only a subset of inputs, whereas k-induction proves that an assertion holds for all reachable states. Our pass rate is therefore lower in part because our acceptance standard is substantially stricter than simulation-based evaluation.

\subsection{Main Results}
Table~\ref{tab:results} reports the results across the six-benchmark suite.

\begin{table}[t]
\caption{Multi-Run Results per Benchmark ($n=5$ runs each)}
\label{tab:results}
\centering
\begin{tabular}{lccccl}
\hline
\textbf{Benchmark} & \textbf{Type} & \textbf{Pass} & \textbf{Iters} & \textbf{Time (s)} & \textbf{Outcome} \\
\hline
alu & comb. & 5/5 & 2.0 & 16.5 $\pm$ 3.0 & proven correct \\
counter & seq. & 0/5 & 10.0 & 47.8 $\pm$ 3.4 & cover vacuity \\
arbiter & comb. & 0/5 & 10.0 & 86.7 $\pm$ 2.8 & spec ambiguity \\
axi\_lite\_slave & seq. & 0/5 & 10.0 & 127.1 $\pm$ 20.9 & multi-property \\
uart\_tx & seq. & 0/5 & 10.0 & 125.3 $\pm$ 18.7 & temporal logic \\
fifo & seq. & 0/3$^{\mathrm{a}}$ & 10.0 & 108.7 $\pm$ 3.5 & temporal logic \\
\hline
\multicolumn{6}{l}{$^{\mathrm{a}}$Three valid runs; two excluded due to an API quota limit,} \\
\multicolumn{6}{l}{not a verification failure.}
\end{tabular}
\end{table}

\begin{figure}[t]
\centering
\includegraphics[width=\columnwidth]{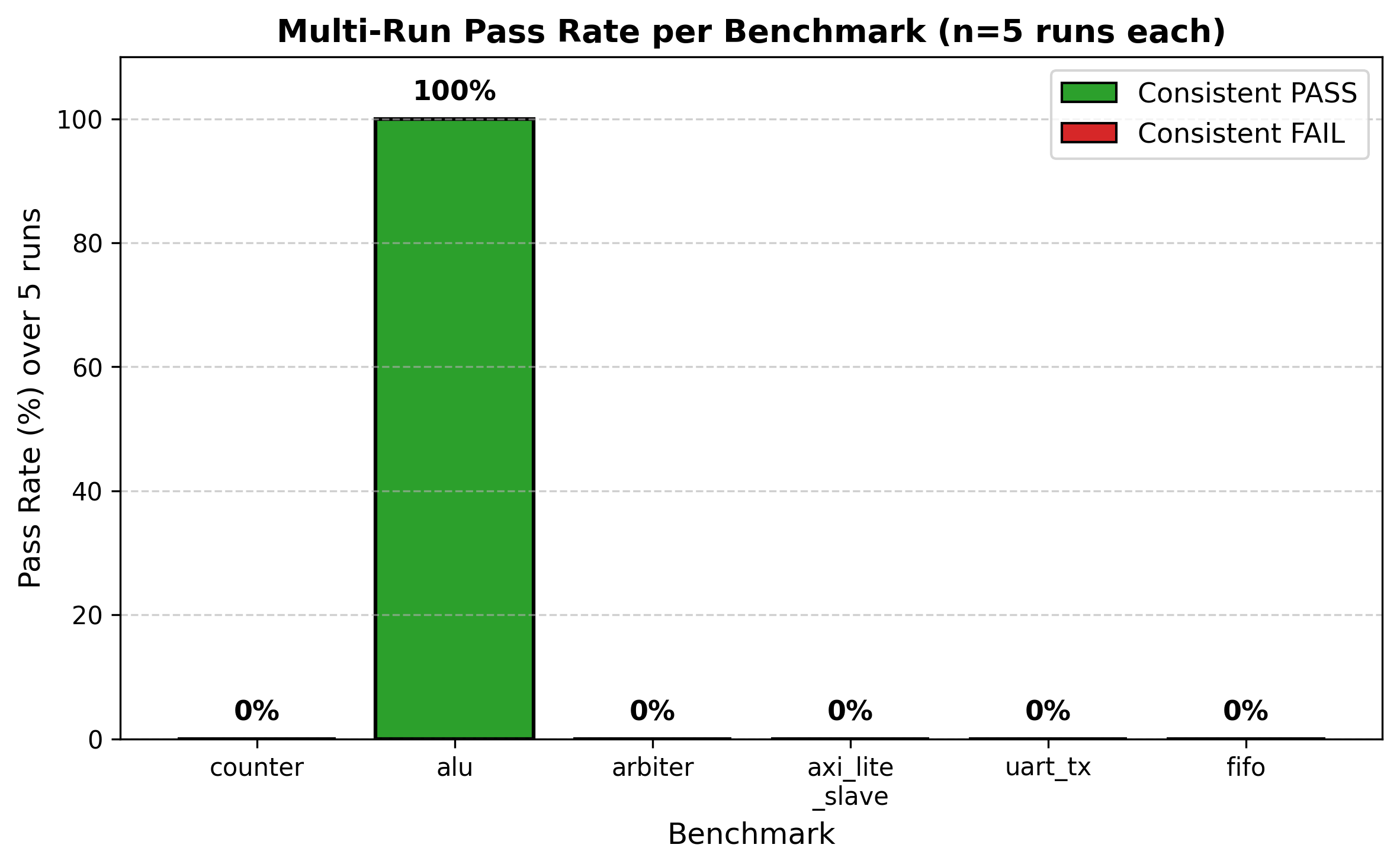}
\caption{Multi-run pass rate per benchmark ($n=5$). For fifo, three valid runs are reflected; two runs were excluded due to an API quota limit.}
\label{fig:passrate}
\end{figure}

Across the suite, one benchmark (alu) is repaired reliably across all five runs, while the remaining benchmarks fail consistently. We report this result honestly: the framework reliably repairs combinational logic in the alu case, and our multi-run protocol exposes distinct, well-characterized failure modes on the others. We analyze these failure causes in detail in Section~\ref{sec:discussion}. Fig.~\ref{fig:passrate} shows the pass rate per benchmark.

\subsection{Case Study: ALU True Positive}
\begin{figure*}[t]
\centering
\includegraphics[width=\columnwidth]{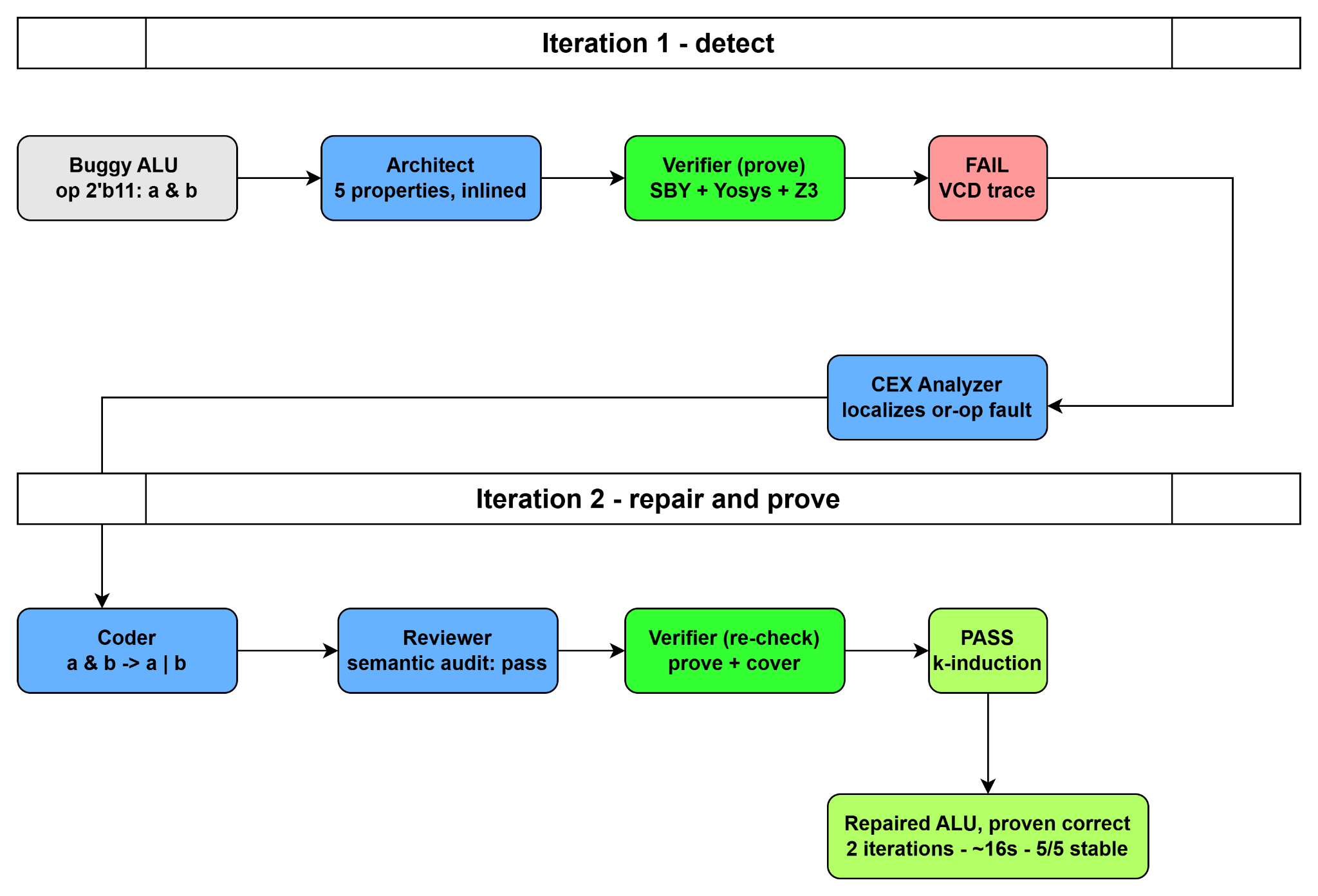}
\caption{End-to-end ALU repair sequence. The pipeline detects the bug via a counterexample in the first iteration and proves the repaired design correct by k-induction in the second.}
\label{fig:alu}
\end{figure*}

The alu module is injected with a single bug: for opcode \texttt{2'b11}, the module computes \texttt{result = a \& b} instead of the correct \texttt{a | b}. The repair proceeds end to end as follows. The Architect generates five properties (one per opcode, plus a zero-flag property) and the Verifier runs a proof, which returns FAIL with a counterexample. The CEX Analyzer parses the VCD trace and localizes the violation to the or-operation. The Coder then applies a single-line fix, and re-verification succeeds by k-induction. The repair changes exactly one character (\texttt{\&} to \texttt{|}), demonstrating that the fix is precise and does not disturb the surrounding structure. The violated property, \texttt{or\_operation\_a}, asserts that \texttt{result} equals \texttt{a | b} for opcode \texttt{2'b11}, allowing the CEX Analyzer to localize the fault precisely. This case is stable across all five runs, converging in two iterations at an average of 16.5 seconds. Fig.~\ref{fig:alu} illustrates the full repair sequence.

\subsection{Iteration and Timing Analysis}
The alu benchmark converges fastest and most stably, requiring two iterations in every run. The five failing benchmarks, by contrast, reach the ten-iteration maximum on every run without converging. Importantly, a high iteration count does not indicate that a repair is close to succeeding; rather, it indicates that the pipeline is stuck---either in a vacuity loop, as with counter, or because the Coder cannot resolve the underlying logic, as with the temporal-logic cases. We also observe that timing variance differs sharply across benchmarks: axi\_lite\_slave ($\pm$20.9~s) and uart\_tx ($\pm$18.7~s) vary far more than the others ($\pm$3~s), suggesting that designs with many properties or deeper state cause the pipeline to struggle in less predictable ways.

\section{Discussion}
\label{sec:discussion}
The framework succeeds on simple combinational logic but fails on the remaining benchmarks through four distinct, well-characterized failure modes. These are not repetitions of a single fault; each arises from an independent cause, and characterizing them is itself a contribution of this work.

\subsection{Failure Mode Taxonomy}
\subsubsection{Bounded-cover vacuity (counter)}
When the counter is verified, every assertion is proved by k-induction, yet the run is still reported as FAIL. The cause is the cover property: reaching the wrap-around state ($\mathrm{count} = 2^{\mathrm{WIDTH}} - 1$, i.e.\ 255 for $\mathrm{WIDTH}=8$) requires at least 256 cycles to become reachable in bounded model checking, and at the default depth it is therefore flagged as vacuous. This exposes a tension between cover-based vacuity checking and designs with deep state. It is a limitation of bounded model checking rather than a repair fault---the repair itself is, in fact, correct.

\subsubsection{Specification ambiguity (arbiter)}
Here the LLM generates properties that cannot be satisfied. The specification describes round-robin behavior, which requires memory, yet the module is declared combinational. The Architect generates fairness and alternation properties that are faithful to the specification but unsatisfiable by pure combinational logic. The underlying issue is a mismatch between the timing implied by the specification and the timing declared by the module. This presents a broader challenge in LLM interpretation of specifications: the model interprets the natural-language intent correctly, but that intent contradicts the declared hardware constraints.

\subsubsection{Temporal logic bugs (uart\_tx, fifo)}
For these benchmarks, the Coder fails to produce a valid fix within ten iterations. The bugs involve multi-cycle temporal behavior that exceeds the Coder's current reasoning ability. This suggests that while the LLM repairs local combinational faults well (as in the ALU case), multi-cycle temporal reasoning remains a challenge.

\subsubsection{Multi-property pressure (axi\_lite\_slave)}
The runs for this benchmark do not converge and exhibit large timing variance ($\pm$20.9~s). Multiple properties compete simultaneously, so a change that satisfies one property may break another, and the pipeline struggles differently on each run. This indicates that repair capability degrades as the number of simultaneous constraints grows---a scalability limitation.

\subsection{What Works}
The ALU case demonstrates the central claim of feasibility: an LLM, combined with open-source formal verification and counterexample feedback, can detect and repair a real bug with a mathematical proof of correctness. The strengths are clear---local combinational faults are fixed with minimal, precise edits and with fast, stable convergence. Equally useful is the sharp boundary between what works and what does not: it indicates which classes of bugs are well suited to this approach and which require further research.

\subsection{Engineering Insight: The Yosys bind Limitation}
\label{sec:yosysbind}
During development we found that the Yosys 0.61 built-in SystemVerilog frontend silently discards modules referenced through a \texttt{bind} directive, emitting only a generic notice that an unused module has been removed. Because the bound assertion module never reaches the solver, SymbiYosys returns a trivial PASS regardless of whether the RTL is correct---a false positive. We confirmed this through a discriminating probe: identical buggy RTL with identical assertions produced a PASS via the \texttt{bind} path but a FAIL with a counterexample via inline injection. We resolve the issue by injecting assertions directly into the design under test rather than binding them. This is a practical note for the open-source EDA community and stands independently of our main results.

\subsection{Threats to Validity}
Several factors limit the generality of our conclusions. The benchmark suite is small (six modules), so the results are preliminary. Each benchmark is injected with a single, manually introduced bug, which may not represent the distribution of bugs that arise in practice or in LLM-generated code. All experiments use a single language model (GPT-4o). Together, these factors mean that our study is best read as a feasibility study rather than a large-scale evaluation.

\subsection{Future Work}
Several directions follow from these findings. Adaptive cover depth for deep-state designs could address the bounded-cover vacuity mode, and refining vacuity detection to distinguish an unreachable cover from a genuinely vacuous assertion would be a more principled solution. A property--timing consistency check could catch specification ambiguity before verification. A stronger Coder, better equipped for multi-cycle temporal reasoning, would target the temporal-logic failures. Finally, expanding the benchmark suite and testing multiple bug types per module are essential next steps.

Taken together, these results position the framework as a feasible but early-stage approach: it establishes that open-source, LLM-driven formal repair is possible, while clearly delineating the technical barriers that must be overcome to make it robust.

\section{Related Work}
\textbf{LLMs for RTL generation.} A growing body of work applies large language models to hardware design, particularly to generating Verilog from natural-language specifications. VeriGen fine-tunes an LLM specifically for Verilog code generation \cite{verigen}, while benchmarks such as VerilogEval \cite{verilogeval} and RTLLM \cite{rtllm} provide standardized problem sets for evaluating LLM-generated RTL. Notably, RTLLM evaluates generated designs against progressive syntax, functionality, and design-quality goals, with functionality assessed through testbench-based checking \cite{rtllm}. Across this line of work, correctness is established by simulation or testing, which exercises only a subset of the input space and therefore cannot provide formal guarantees.

\textbf{LLM-based RTL repair with feedback.} Closer to our work, RTLFixer combines compiler feedback with retrieval-augmented generation and ReAct-style iterative prompting to repair Verilog code, resolving 98.5\% of compilation errors in its debugging dataset \cite{rtlfixer}. However, RTLFixer targets syntax errors and relies on compiler logs and simulation as its feedback signal; its authors report that extending the approach to logic errors detected through simulation yields limited improvement. In contrast, our framework targets functional logic bugs and uses formal counterexamples---concrete waveform traces produced by an SMT solver---as the repair signal, with correctness established by k-induction proof rather than test passing.

\textbf{Positioning.} Existing approaches thus either generate RTL and validate it by simulation, or repair RTL using compiler and simulation feedback. To the best of our knowledge, the combination of LLM-driven repair, formal counterexample guidance, and an entirely open-source verification toolchain remains largely unexplored; this intersection is the focus of our work.

\section{Conclusion}
We presented a multi-agent pipeline that combines a large language model with open-source formal methods for counterexample-guided RTL repair. Through an ALU case study, we showed that the framework can detect a real functional bug, repair it, and prove the result correct by k-induction, establishing the feasibility of the approach. Across a six-benchmark suite, however, only one design was repaired reliably, and we characterized four distinct failure modes---bounded-cover vacuity, specification ambiguity, temporal-logic bugs, and multi-property pressure---that mark the current limits of the method. This boundary between tractable and intractable cases, together with a practical note on a limitation of the Yosys \texttt{bind} directive relevant to the open-source formal verification community, is a central contribution of this work, showing that combining open-source formal verification with an LLM is feasible in principle. Future directions include refining vacuity detection for deep-state designs, strengthening the model's multi-cycle temporal reasoning, and expanding the benchmark suite to bug types beyond those studied here. By building entirely on freely available tools, we hope this work lowers the barrier to combining language models with formal methods for reliable hardware design. 

\textbf{Availability.} The framework, benchmarks, and the raw
multi-run logs underlying Table~\ref{tab:results} are available at
\url{https://github.com/trunghafromvietnam/rtl-repair-framework}.

\end{document}